# The origin and early evolution of life in chemical complexity space

David A. Baum, Department of Botany and the Wisconsin Institute for Discovery, University of Wisconsin, Madison WI 53706, USA; dbaum@wisc.edu

**Abstract**

Life can be viewed as a localized chemical system that sits on, or in the basin of attraction of, a metastable dynamical attractor state that remains out of equilibrium with the environment. Such a view of life allows that new living states can arise through chance changes in local chemical concentration ("mutations") that move points in space into the basin of attraction of a life state – the attractor being an autocatalytic sets whose essential ("keystone") species are produced at a higher rate than they are lost to the environment by diffusion, such that growth in expected. This conception of life yields several new insights and conjectures. (1) This framework suggests that the first new life states to arise are likely at interfaces where the rate of diffusion of keystone species is tied to a low-diffusion regime, while precursors and waste products diffuse at a higher rate. (2) There are reasons to expect that once the first life state arises, most likely on a mineral surface, additional mutations will generate derived life states with which the original state will compete. (3) I propose that in the resulting adaptive process there is a general tendency for higher complexity life states (i.e., ones that are further from being at equilibrium with the environment) to dominate a given mineral surface. (4) The framework suggests a simple and predictable path by which cells evolve and provides pointers on why such cells are likely to acquire particulate inheritance. Overall, the dynamical systems theoretical framework developed provides an integrated view of the origin and early evolution of life and supports novel empirical approaches.

All cellular life that we know of is organized in cells that depend upon biopolymers to transfer information across generations and to fulfill the essential chemical functions of life, namely survival, reproduction, and evolution. However, just because our experience is limited to such life forms does not mean that this is the only way of being alive. For one thing, we do not know if there might be other kinds of entities that can evolve adaptively, justifying them being called alive, that do not form discrete cells and/or do not contain informational biopolymers. For another, if we wed ourselves too strongly to the notion that living systems must be composed of discretely individuated genetically-encoded units in order to evolve adaptively, then we close off the possibility that some or all of these attributes could have arisen via a prior adaptive process (Baum, 2015). Thus, constructive research into the origin of life needs a framework in which modern cellular life represents one of many possible ways of being alive. In other words, we need a general conceptual framework for thinking about life and its evolution within which cellular life is a special case.

In this paper I develop such a generic description of life and using it to explain how life can achieve the high level of complexity we see today. After all, one thing that is strikingly weird about life is its apparent ability to overcome the second law of thermodynamics and become more complexly ordered over the eons. To accomplish this ambitious goal, what is needed is a conceptual framework in which tendencies (if present) can readily be understood in relation to entropy.  Here, I propose that living systems correspond to metastable attractor states in a chemical dynamical system and then use this framework to explore the earliest emergence of life from non-life, its path towards complexification, and the causes and consequences of cellular encapsulation.

Chemical complexity space

Stipulating that life is a spatially localized, chemical phenomenon that uses fluxes of energy and chemical foods to overcome entropic decay and/or to grow, it is ideal to visualize an abstract space that is conditioned upon the expected concentrations of chemical species in the absence of life-like processes. We will achieve this by considering the concentrations of all chemical species in a small but finite area of space, a pixel[1], *relative to these same chemicals' concentrations in the environment.*

Suppose that we had complete knowledge of the kinetics of all possible chemical reactions. In such a case, given the concentrations of all possible chemical species ($X_1$, $X_2$, $X_3$…$X_N$) in a single pixel, $p$, surrounded by pixels that are at equilibrium with the environment, at a certain time, $t$, we would be able to calculate the expected changes in concentration of all species in $p$: $d[X_{1-N}]_p/dt$.

Imagine an environment that experiences a constant flux of a set of environmentally provided chemicals, at concentrations, $[X_{1-N}]_e$. The current state of a certain pixel in this environment can be summarized by placing it in a multidimensional space in which each axis is the difference between the chemical concentration in $p$, $[X]_p$, and in the environmental flux, $[X]_e$, weighted by the internal chemical energy of that species, $U_x$. Thus, the coordinate of $p$ in this multidimensional space is: $U_{x1}([X_1]_p - [X_1]_e)$, $U_{x2}([X_2]_p - [X_2]_e)$, $U_{x3}([X_3]_p - [X_3]_e)$… $U_{xN}([X_N]_p - [X_N]_e)$.

By focusing on concentration differences between the pixel and the environment, this abstract space allows us to capture the notion that a living pixel is one that is far from being in equilibrium with the environment that it lives in. Weighting by energy content is needed, however, to capture the idea that we are not equally concerned with the local concentration of all chemical species, but specifically of high-energy species that would be unlikely to be present at high concentrations by chance. Finding a pixel of concentrated water in an environment rich in dihydrogen and dioxygen would hardly be remarkable. But finding peptides in an environment through which ammonia and carbon dioxide flow might be suggestive of life-like activity. To capture this factor, I propose weighting the coordinate system such that the position of a pixel on an axis is proportional to the probability that this chemical species would be formed and then concentrated to such an extent by chance alone. I recognize that the summation over the bond energies of chemical species, its internal energy, is not a practical (i.e., measurable) attribute in most cases, but I will argue that it is sufficient for the conceptual exploration we will be conducting here.

---

[1] A pixel is offered as an aid to conceptualization but is not an essential component of the underlying theory, which could be cast in continuous space (at least down to the level of individual molecules). To serve its conceptual role, imagine a pixel as an area or volume that is large enough that all chemical species that are relevant for predicting its expected dynamical behavior have a high probability of being present (at their current concentration), yet small enough that diffusion renders it well mixed.

The concentrations of some chemical species at $p$ may be lower than in the environmental flux because the chemical reactions occurring at $p$ have used more of that chemical than they produce. This means that on some axes, $p$ may be to the left of the origin. However, given that life's novelty entails maintaining certain chemicals at high local concentration, my illustrative figures will focus only on those dimensions for which $p$ has positive values. Viewing the environmental flux as food, chemicals in $p$ whose current concentrations exceed that in the flux, the non-food chemicals[2], must have been produced in the vicinity of $p$. Figure 1 represents an idealized sketch of this space, compressed onto two arbitrary axes.

The origin of this coordinate system corresponds to a pixel that is in complete equilibrium with the environmental flux, which can be viewed as the point of maximum entropy. Even in an idealized environment with constant flux, individual pixels will not all sit exactly on the origin. Microscale stochasticity of molecular systems ensures that any given pixel is likely to deviate from the environmental mean. However, if all that is at play is chance, then the probability of a pixel being at a certain coordinate will correlate with the distance of that coordinate to the origin[3]: points distant from the origin cannot be found by chance but require some long-term, adaptive process. As a result, taking the view that the most interesting kind of complexity for thinking about the origin of life is the improbability of spontaneous appearance (Baum, 2015), then the coordinate system can be interpreted as a space of *chemical complexity*. The Euclidean distance between a pixel and the origin is, therefore, a rough measure of that pixel's chemical complexity (CC):

$$CC_p = \sqrt{(U_{X1}([X1]p - [X1]e))^2 + (U_{X2}([X2]p - [X2]e))^2 + (U_{X3}([X3]p - [X3]e))^2 + \cdots + (U_{XN}([XN]p - [XN]e))^2} \quad (1)$$

---

[2] It is a priori improbable that chemicals produced in $p$ will already be present in the environmental flux, if only because Le Chatelier's principle would disfavor reactions producing readily available products, so I will label all chemical species present at above-background concentrations as "non-food" chemicals.

[3] This is an oversimplification. The relative importance of concentration versus internal energy as predictors of "the probability of spontaneous appearance" will vary depending on whether a chemical is concentrated in a pixel as a result of local synthesis or of stochastic variations in space. In the former case internal energy could greatly affect this probability, but in the latter, concentration alone ought to be sufficient to predict the probability of a certain jump in concentration.

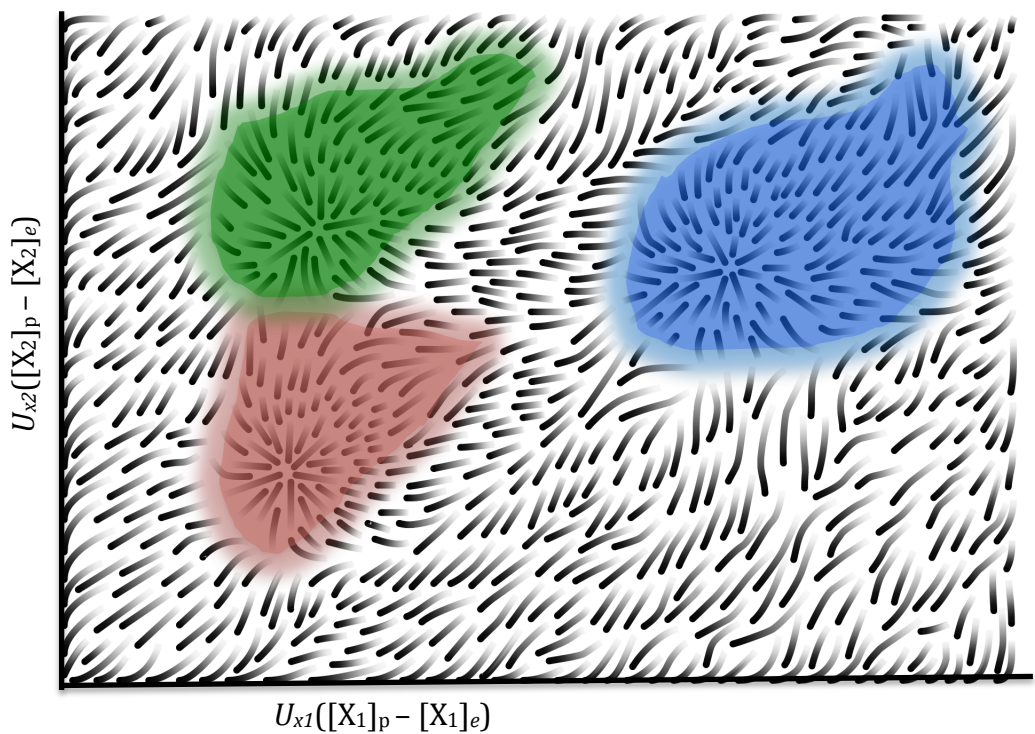

Fig. 1. Illustration of chemical complexity (CC) space, where each point corresponds to the difference in concentration of all chemical species between a specified pixel, $[X_i]_p$, and the environment, $[X_i]_e$, weighted by the internal energy of that chemical species $U_i$. CC space has as many axes as there are chemical species, but just two axes are shown for illustrative purposes. The streaks depict the expected change in chemical compositions of a pixel as a consequence of chemical reactions occurring in the pixel and in the absence of noise or interactions with other pixels. The shaded areas correspond to the zones of attraction around three different metastable attractors, here interpreted as life states.

The dynamics of a pixel over time in CC space will depend on four factors: (a) reactions occurring among the chemicals currently at *p*, as governed by their current concentrations and the kinetic rules of the full dynamical system, (b) the diffusive exchange of chemicals between *p* and the bulk environment, (c) the diffusive exchange of chemicals between *p* and adjacent pixels, and (d) chance fluctuations. The first three factors can be represented as expected vectors of change in the position of *p*. The first factor yields a vector that could point in any direction in any axis depending on the kinetics of all reactions in the dynamical system. The second factor provides a constant vector pointing towards the origin, whose magnitude is proportional to the distance from the origin, scaled by the rate of diffusion. The third factor yields a vector pointing towards the concentrations in the adjacent pixel(s), scaled by diffusion. The fourth factor, chance, yields expected changes on each axis that are exponentially distributed with a mean of zero.

Considering a focal pixel, and ignoring adjacent pixels (for now), a few generalities are worth highlighting. In the absence of chemical reactions occurring in *p*, *p* will move towards the origin, the point of minimum complexity. This aligns with intuition: unless active processes are at play, chemical systems will equilibrate with the environment.

However, *p's* movement towards the origin in CC space can be deflected or reversed by chemical reactions occurring in *p*.

CC space as a whole can be imagined as a field of vectors corresponding to the expected changes in the CC of an isolated pixel. In such a space a preponderance of vectors will point towards the origin (Fig. 1). An alternative way to visualize this is as a landscape in which vectors due to equilibration with the environment correspond to the expected gravitational pull on a ball, corresponding to a pixel's chemical state. As shown in Fig. 2, every point in CC space has higher altitude than the origin, meaning that the overall tendency of pixels is to fall towards the origin.

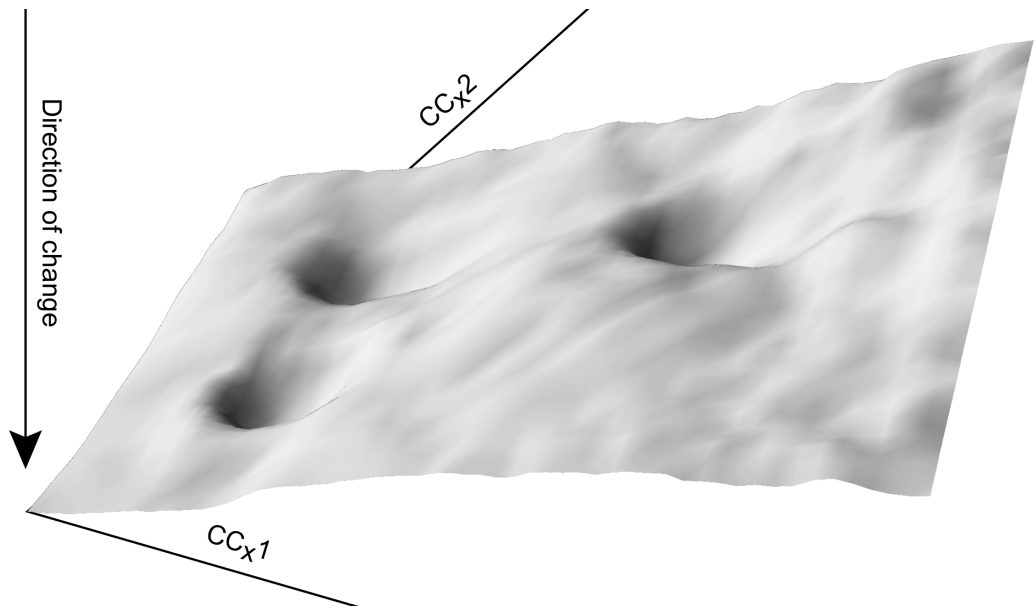

Fig. 2. Illustration of chemical complexity (CC) space with expected changed in chemical concentration of a pixel depicted in terms of the expected position of a ball responding to gravity. The three wells correspond to three metastable equilibria and their zones of attraction (as in Fig. 1)

We know that within the real (but incompletely charted) chemical dynamical system, some pixels, most notably those corresponding to living cells, persist far from environmental equilibrium. The best way to understand this is to equate a living state with a metastable attractor in CC space[4]. In reality these attractors could be of various kinds (fixed point, limit cycle, strange, etc.) but, for simplicity, let us visualize each metastable attractor as a point surrounded by a basin of attraction. By definition, a pixel in the basin of attraction will tend to move towards the attractor and then remain there until one of three things happens: the physical environment changes (altering the kinetics of chemical reactions within *p*), the chemical composition of the environmental flux changes, or a stochastic event knocks the pixel out of the basin of attraction.

---

[4]Pross (2005; 2012) called the property of being in such a state, "dynamic-kinetic stability," which helpfully captures the kind of stability entailed.

How can a pixel remain at an attractor and avoid falling to equilibrium? This is only possible if the reactions in *p* use energy to prevent entropic decay (as famously pointed-out by Schrödinger (1945). That is to say, the second law of thermodynamics means that, in order to prevent or delay the fall towards the origin, the net reactions in *p* must be endergonic[5]. Additionally, since all the non-food chemical species in *p* are susceptible to dilution by environmental flux, we can also conclude that if *p* sits at a metastable attractor, all non-food species must be produced by reactions within *p* at a rate equaling their rate of dilution. If the rate of production were lower, *p* would not be sitting directly on an attractor, and would drop towards the origin (which also may be towards the closest attractor if the pixel is in a basin of attraction). If the rate of production is higher than the flux, then *p* will move further away from the origin until it sits on the actual attractor.

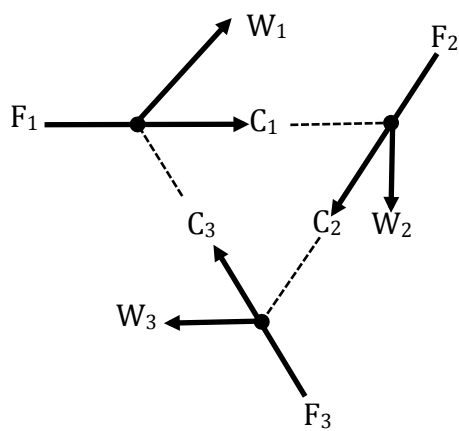

Fig. 3. Example of a simple 3-member autocatalytic set. Three kinds of food chemicals, $F_1$, $F_2$, and $F_3$, are converted to three catalytic species, $C_1$, $C_2$, and $C_3$, respectively, with the simultaneous production of three waste products, $W_1$, $W_2$, and $W_3$, respectively. The chemical conversion of food to catalyst plus waste is represented with solid arrows, while dotted lines indicated catalysis.

As an aid to visualization, Fig. 3 shows a set of chemical reactions whose kinetics might allow them to form a spatially localized metastable state in which the non-food catalytic species, $C_1$, $C_2$, and $C_3$, and non-food waste species, $W_1$, $W_2$, and $W_3$, become stabilized at high local concentration in an environment rich in a replenishing supply of food species, $F_1$, $F_2$, and $F_3$. A pixel having a high concentration of $C_1$, $C_2$, and $C_3$ might sit on a metastable attractor state such that the rate at which these three species form equals the rate at which they are lost to the environment by diffusion. Such a system[6] would contain an ensemble of chemicals that make more of one another over time, meaning that we can refer to the system as either autocatalytic (regardless of whether specific catalysts are present) or self-propagating. Pixels in the basin of attraction of a metastable equilibrium could also be autocatalytic, although the rates of production of each functional chemical would not exactly match the rates of dilution until the attractor point is reached.

A dynamical definition of life

---

[5] It is worth noting that the second law does not place a limit on the amount of chemical energy concentrated in *p* because non-food chemicals can accumulate over time. As a result, the net internal energy concentrated in *p* can (and in general will) exceed that in the environment. This accords with the observation that the energy density of life tends to greatly exceed that of the planet as a whole.
[6] A "system" in this context is a set of chemicals that defines a dynamical state that has some degree of persistence through time.

Within the framework just presented life corresponds to a localized collection of chemical species sitting at a metastable equilibrium in CC space or in the basin of attraction around such a metastable equilibrium. In contrast, pixels that sit at the origin are dead, whereas pixels that sit neither at the origin (global death) nor in the gravitational pull of a metastable equilibrium (alive) are on a trajectory towards the origin, i.e., death. We might consider them probabilistically dead, in the sense that unless some perturbation moves them back into the life state they will deviate further and further from it, with ever-decreasing probability of autonomously returning to life[7].

It is worth asking at this point whether the notion of life as a localized chemical system in the basin of attraction of a metastable attractor captures the important features of more conventional definitions of life. First it is generally held that being alive entails an ability to grow. To remain at an attractor state, the chemicals in *p* must produce enough of the self-same chemicals that are enriched at *p* to compensate for those lost by diffusive exchange with the environmental flux. Since diffusion is an inherently spatial phenomenon, we can see that a pixel adjacent to *p*, *p'*, will tend to sit close to, and on the downhill side, of *p* in CC space (Fig. 4). If *p'* sits within the zone of attraction around the living the living state occupied by *p*, it will move to the same state as *p*. The expansion of the life state from *p* to *p+p'* is growth. Furthermore, assuming environmental constancy, *p'* will convert its adjacent pixels, *p''* to the life state, and so forth. Thus, my definition captures one of the most commonly emphasized features of life, namely the capacity for open-ended growth in a permissive environment[8].

What about another commonly emphasized feature of "life," the capacity for evolution? At small physical scales, chemical kinetics is shaped by molecular stochasticity as much as by mass action kinetics. As a result, a pixel sitting at one position in CC space always has a finite probability of jumping to a new position that would not have been predicted based on the general dynamical model. Given that such jumps involve changes in chemical concentration, and smaller changes in concentration are more likely than large ones, it follows that the expected position of *p* after undergoing such a stochastic "mutation" is centered on its current position and declines with distance in CC space (but see footnote 3). Mutational jumps have the potential to instantaneously move *p* from near one attractor state into the basin of attraction of another. As illustrated by Wynveen et al. (2014) many chemical networks have the potential to manifest multiple metastable equilibria. This means that a mutation can result in *p* adopting a new life state, with an expectation that *p* will remain in the derived state until perturbed by a change in the environment or by a further mutation. Thus, it is in principle always possible (if in some cases vanishingly improbable) for a pixel to evolve from one metastable life state to another.

---

[7] It should be born in mind that loss of life from a pixel could correspond to movement of life from that pixel into an adjacent pixel. Likewise, returning to the life state could correspond to movement of a life state into this pixel rather than reincarnation. It is only after life acquired bounding membranes that it became possible to distinguish motility from growth and allow for clear notions of death or reincarnation.

[8] If it does not grow, then we can conclude that the state in *p* is not at a metastable equilibrium in a spatially explicit context: it is expected to shrink to non-existence rather than grow. Thus, ignoring the improbable case in which concentrations around a spatially-localized chemical state decline in such as way that the adjacent pixel sits *exactly* on the edge of the zone of attraction, being alive implies the capacity to grow.

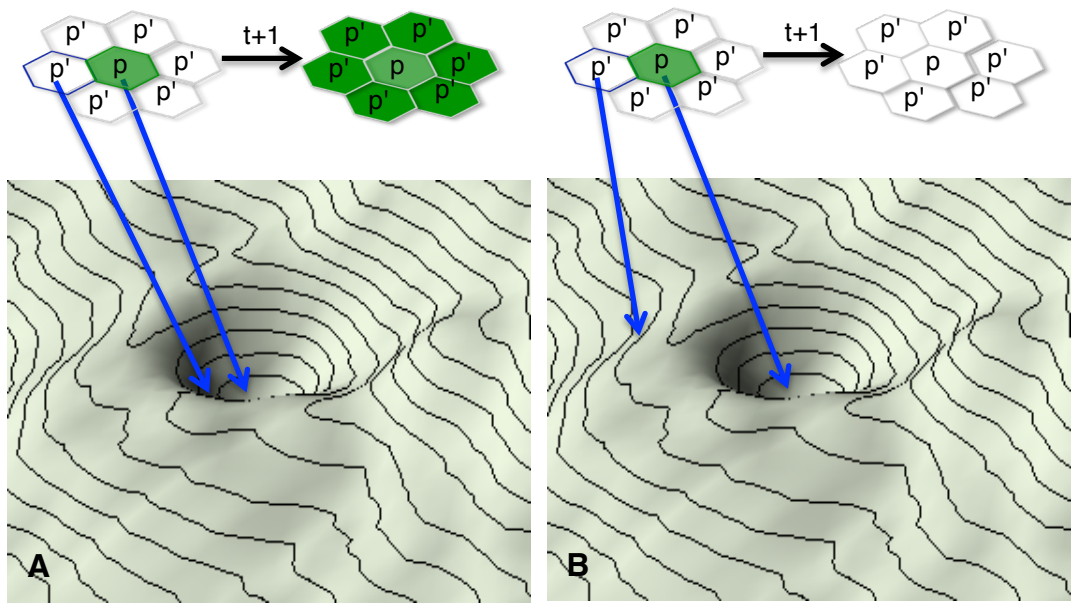

Fig. 4. The viability of a life state in an explicit spatial context depends on the state expected for a pixel, p', adjacent to a pixel that sites on the attractor of a life state. A. If p' receives sufficient influx of chemicals from p to situate p' in the zone of attraction, the life state will grow from p to encompass all adjacent pixels. In this case the life state is viable. B. If p' sits outside of the zone of attraction, for example because dilution of chemicals enriched at p is too rapid, p' will not be converted to the life state. In this case the life state is not viable and the area in the life state will shrink over time.

The movement in CC space just described might be discounted as "just" change since there is no adaptive process entailed. However, as discussed more fully below, adaptive evolution would be in effect if life states differ in their expected longevity such that pixels in robust chemical states should become more abundant than pixels in less stable life states. For example, all things being equal, life states with large basins of attraction will tend to persist for longer because they are less likely to exited by chance concentration changes. Additionally, the ways that life states interact could cause some to win at the expense of others. For example, if *p* finds itself adjacent to a pixel, *p'*, that occupies a different life state, properties of the two states could result in the *p* state taking over *p'* or vice versa. If this transpires, a selective process would result in certain chemical states, which may be classified as fitter in the specified environment, tending to outcompete less fit alternatives. These phenomena will be discussed in more detail later. For now, all that needs to be noted is that life as I have defined it has the potential for both growth and for adaptive evolution.

The reader might object that some chemical states that fit this definition deviate markedly from life, as conventionally defined. For example, a pixel inside a candle flame[9] receives a flux of long-chain hydrocarbons delivered by the wick and oxygen delivered by convection, which result in a local enrichment of a different set of species in the flame, for example shorter-chain hydrocarbons, carbon monoxide, and carbon dioxide. However, while a flame is alive by my definition and capable of growth, for example, along a wick or to a second wick, I will argue that it is not an interesting life state because it probably cannot evolve. A flame can change over time as a result of changes in the environment or in the flux of fuel and oxygen, but evolution, per se, is not possible because the chemical species that constitute a flame all cycle through the system at about the same rate (on the same time frame as energy dissipation). As a result, a flame lacks a class of chemical species that can store "memories" of past conditions. In contrast, a living chemical system has a set of non-food species whose residence time, though finite, is much greater than the rate at which food is digested and waste released. The identity and concentration of these non-food species is influence not just by the current environment, but also by the recent history. It is these non-food species, therefore, that cause the system to tend to remain on an attractor unless perturbed into another state (at which it will then tend to remain). Thus, I will posit that only a subset of dissipative structures in nature, namely those that entail some longer lived chemical species, provide a basis for complexification through adaptive evolution.

---

[9] It is worth noting that combustion requires a liquid or solid fuel that is greatly out of redox equilibrium with the local environment. Such conditions on Earth are mainly the result of life – with reduced organic matter serving as the fuel and oxygen as reactant.

The spontaneous appearance of surface-associated life

Before going on to explore adaptive evolutionary processes, it will be helpful to develop a sense for where evolvable life states are most likely to arise in the first place. A new self-propagating chemical state needs both food, which is to say the components that will make the chemical species in the autocatalytic set, and energy to drive a set of reactions that are, in net, endergonic. However, neither factor is likely to have been infrequent on the early Earth, on other planetary bodies, or even in many natural and human-made Earth environments today. Even if we restrict our attention to organic life, we have known since the Miller-Urey experiments and investigations of cometary material that potential food species form spontaneously in many situations. Furthermore, there are many natural energy sources in addition to high-energy bonds in the organic foodstuffs themselves, most obviously light and redox or pH disequilibria maintained by radioactively-driven geochemical processes. Under what circumstances can an environment that experiences flux in food chemicals, with or without additional source of chemical energy, transition to an evolvable living state?

The main factor determining whether a life state is viable in a spatially explicit model is diffusion, and particularly different rates of diffusion for food and non-food chemical species. Diffusion of food chemicals into the system is needed for self-propagation, with higher rates of diffusion supporting potentially higher rates of production of non-food chemicals. Likewise, diffusion of waste species out of the system will often be needed to sustain high enough rates of autocatalysis. At the same time, it is essential that diffusion not result in the loss of those non-food chemical that are needed for self-propagation, the *keystone species*[10]. A potential life state is not viable if its keystone chemicals (e.g., catalysts

---

[10] This seems like an appropriate re-tooling of a concept from ecosystem ecology (Paine 1969; Mills et al. 1993). In both cases the species are defined based on the counterfactual: if

$C_1$, $C_2$, and $C_3$ in Fig. 3) diffuse away from one another faster than they are produced. While formal analysis is needed, I will conjecture that life cannot easily arise when these two classes of diffusion rate are equal. It is possible that the keystone chemical species might happen to have a lower diffusion rate than food chemicals, for example by being larger molecules. However, it seems a priori easier to achieve the life state if the keystone chemical species are configured such that they have a low rate of diffusing away from each other while still having access to rapidly diffusing food species.

One setting that seems especially conducive to differential diffusion is at the interface between two different phases: solid-liquid, solid-gas, or liquid-gas. At such an interface an autocatalytic set of chemicals could persist and grow if the diffusion of the keystone species were governed by the less diffusive phase while food was provided by the more diffusive phase. Indeed, the origin of life field has been focused on the solid-liquid interface at least since Wächtershäuser (1988). This setting seems well suited to the origin of life because we know that in realistic geological settings the liquid phase, for example, an ocean or volcanic pool, can become enriched in diverse chemical species, some of which will be adsorbed selectively onto specific kinds of mineral surface. Additionally, some mineral or metallic surfaces can serve as keystone species themselves, for example by serving as direct catalysts of one or more reactions in an autocatalytic set.

Under this model, life could begin when a set of keystone species become attached, by chance, to the same pixel of a mineral surface. For example, in the hypothetical autocatalytic set shown in Fig. 3, if the keystone catalysts, $C_1$, $C_2$, and $C_3$, tended to become immobilized on a particular mineral surface, while the food and waste species did not, a life state might nucleate and then grow over the surface as more catalyst molecules are formed retained in close proximity by adsorption onto adjacent surface. Provided diffusion of food and waste is fast enough relative to the rate at which keystone species are lost from the surface, one-ended growth of the life state over the surface would be possible. Consequently mineral surfaces seem to be prime sites for *de novo* life states to nucleate spontaneously and then propagate themselves. Although many of the principles I will discuss below do not formally depend on a solid-liquid interface, for linguistic simplicity I will hereafter assume that new life states emerge adsorbed onto solid surfaces interacting with a food-rich liquid phase.

Mutations in CC Space

At this point we may imagine the abstract CC landscape, with many potential metastable attractor states, one of which has become occupied by the spontaneous formation of an autocatalytic system in one pixel on a mineral surface bathed in a replenishing, food-containing liquid. From such a starting point, how can evolution explore CC space and, most particularly, how can new attractor states uphill from the initial life state, that is further from the environmental equilibrium, be found?

---

the species were not present at all, the dynamical state (life or a specified ecosystem composition) would not persist.

As already indicated, some exploration of CC space can be achieved by chemical mutations. Thus far I have emphasized changes in concentration caused by sampling error at small spatial scales, which can move a pixel in any direction in CC space, with the probability of a distant move being lower than that for a near move. Now we need to add another kind of mutation: rare chemical reactions, such as ones requiring the simultaneous interaction of multiple, low-concentration reactants. These rare chemical reactions can introduce a new chemical species into a system[11], which amounts to opening a new dimension in CC space.

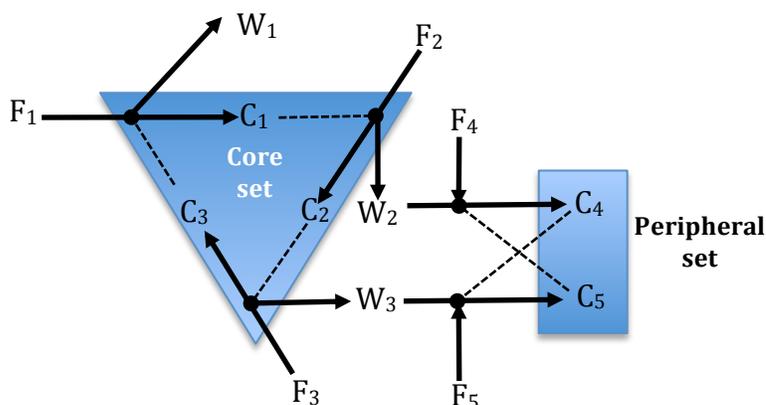

Fig. 5. Example of how rare chemical reactions, such as the localized production of $C_4$ and $C_5$ by the slow spontaneous reaction of waste products from the original catalytic set with low-abundance food species, could expand the self-propagating system, resulting in movement of a pixel to a new metastable state (with high concentrations of $C_4$ and $C_5$). It is supposed that the newly formed non-food species $C_4$ and $C_5$ cross-catalyze one another formation and tend to attach to the same mineral surface as $C_1$, $C_s$ and $C_3$. The chemical conversion of food in catalyst plus waste is represented with solid arrows, while dotted lines indicated catalysis.

Such chemical novelty is illustrated in Fig. 5, which represents an extension to the autocatalytic core shown in Fig. 3. In this case, if two new chemical species $C_4$ and $C_5$ formed spontaneously on the surface, for example by a rare reaction between two waste products from the original set and two further food compounds, then these two non-food species could become added to the original set representing a shift to a new metastable equilibrium that entails non-zero concentrations of $C_4$ and $C_5$.

The chemical mutational processes just described would be random with respect to the dimensions of CC space. Depending on whether a new species' production causes compensatory changes in the concentration of other chemicals, the new life state could have higher or lower chemical complexity than the original state – jumps away from the origin are about as likely as ones towards the origin. This means that chemical mutations have the potential to continually generate higher complexity states, some of which might be favored during adaptive evolution (see Baum 2015, Fig. 1).

---

[11] It is perhaps worth mentioning that new chemical species continue to be added to cellular life, as seen most prominently in the evolution of new secondary metabolites by plants and microbes. Additionally, it should be borne in mind that a mutation in a DNA molecule can be envisaged as the loss of one chemical species (present as a single copy per cell) and the appearance of a different one. That is to say, genetic mutations are just a special case of chemical mutations in general.

In addition to chemical mutations, CC space can also be explored by changes in environmental fluxes or in the physical environment. These changes can push living systems into new states, some of which could be self-propagating. Furthermore, it is possible that a sudden change back to the prior environmental parameters could result in a living system being left in the zone of attraction of a different attractor state than previously. As with the conventional evolutionary history of cellular life, such historically contingent environmental buffeting is unpredictable and not readily explained by general laws, but no less real or important for this unpredictability.

<u>Evolution to higher dynamic chemical complexity</u>

Each potential life state, whether or not it is occupied, corresponds to a dip or divot in CC space; which is to say an attractor state surrounded by a basin of attraction. The complexity of those potential life states, namely the minimum Euclidean distance from the origin to a point within the basin, varies greatly. The first life state occupied in an evolutionary sequence arises via chance changes in local concentrations meaning that it must be of low complexity. Thus life will always start close to the origin. The fact that even the simplest known living systems are far too complex to have arisen spontaneously, implies that life began at lower levels of complexity and, over time, leap-frogged through a series of life-states that were sequentially further and further from the origin. How and why did this happened?

Some evolutionary biologists, most famously Stephen Jay Gould believed that the fact that life must begin at the left wall of a complexity graph is sufficient to create an illusion that there is a drive to complexity (Gould, 1988). In other words there is no bias towards higher complexity just a constraint to begin simple. This case might be defensible when it comes to cellular life, but if you believe (as I do) that a cell is much too complex to arise without the prior existence of a living and evolving system (Baum, 2015) then the lack of life systems sitting in the chasm between cells and the origin of CC space suggests that the left-wall artifact is insufficient to explain the observed pattern. This is, I think, why many scientists (e.g., Krakauer, 2011) take it as a given that the remarkably intricate life forms around us today point to some intrinsic evolutionary drive to higher complexity[12]. Can we use the CC framework to evaluate whether adaptive evolutionary processes acting prior to cellularization (i.e., on surface-associated systems) might favor high CC relative to low CC life states?

First we need to consider if complexity itself might tend to enhance the long-term persistence of self-propagating systems living on mineral surfaces. Self-propagating systems can always be bounced out of the life stated by chance changes in the concentrations of chemical species. The distance that a stochastic mutation is likely to move a pixel is dependent on current concentrations: low abundance species are expected to change proportional concentration more than high abundance species[13]. This means that, assuming all other factors are equivalent (i.e., the size of the zone of attraction, energy content, and the distribution of concentrations over species) higher complexity states, which have higher mean concentrations, should be more robust to mutation than lower complexity states. This mechanism is similar to that analyzed more formally by England, (2015). He suggested that, when the environment is rich in potential energy sources that are difficult to access, life states that can exploit this energy gain stability because they dissipate energy locally, which makes it more difficult for a random process to take them out of the life state.

---

[12] This is not to deny that simpler life states can sometimes be favored, as seen for example when lineage move from a diverse and uncertain environment into a much more stable and predictable one, as happens during the evolution of endosymbionts. Rather, the claim is that the net flow is towards higher complexity not withstanding some eddies that yield lower complexity life states.

While the foregoing argument is attractive, many other factors could swamp the mutational robustness advantage of high complexity states. For a start, once a life state becomes nucleated and grows over a surface, it acquires an internal correction system that should give it resistance to death-by-mutation. If one pixel of a larger living system were knocked out of the life state and began to equilibrate with the environment, growth from adjacent pixels would tend to return the dead pixel back to the life state[14]. Indeed, one of the beauties of the surface metabolism model for early life (e.g., Wächtershäuser, 2007) is that large enough surface systems in the life state are resistant to lethal perturbations whilst still allowing many opportunities for chemical mutations into new life states that can then propagate through the whole system. As a result, differences in growth rate among living states, which would not obviously correlate with complexity[15], could be the main evolutionary driver, casting doubt on my preceding claim that mutational stability is sufficient to explain complexification of surface associated life states

A further important driver of complexity change might reside in the outcomes of interactions of adjacent pixels that are in different life states. When a pair of unbounded, growing life states come into contact (Fig. 6) there are three possible outcomes: (1) Annihilation: The life states cancel each other out, meaning that the pixels are both moved out of the life state; (2) Coexistence: The states merge, meaning that both chemical states coexist in each pixel – they grow through one other; (3) Replacement: One state wins, meaning that it grows through the other state converting pixels from the losing to the winning state.

---

[13] Specifically the effect should scale with one over the square root of the number of molecules.

[14] In addition to conferring mutation resistance, spatial extent also protects living systems from parasitic chemical species, species that catalyze their further production in a pixel (and maybe adjacent pixels) at the expense of one or more keystone species. As shown by Boerlijst & Hogeweg (1995) and Könnyű et al. (2008), the spatial structuring of self-propagating systems provides some degree of parasite resistance. Indeed, in spatially explicit models, parasites have a pretty hard time killing a host and parasites may even yield unexpected evolutionary benefits (Virgo et al. 2013).

[15] The major determinant of growth rate is likely not the rate of production of keystone species, as one might have thought, but the keystone species' rate of diffusion over the surface.

Annihilation is an expected outcome when the systems each depend on a certain flux of the same food chemical, which is insufficient to support either autocatalytic set above its rate of diffusive loss. Alternatively, interactions among non-food chemicals could produce poisonous or parasitic products. However, since annihilation removes two life-states, it can play no role in increasing the complexity of the life states that do exist, making it irrelevant to the question at hand.

Coexistence increases the number of chemical species that have elevated concentration, thereby raising complexity further from the environmental equilibrium point. This outcome is facilitated by cases in which the non-food chemicals enriched in the two autocatalytic sets do not compete for food or attachment points on the surface and do not cross-react to produce poisons. Coexistence could be passive, in the sense that neither products or reactants are shared, such that the original autocatalytic systems neither compete nor cooperate with one another. Sometimes, however, coexistence might entail cooperation, as when the waste products of one are used and/or degraded passively by the other (see Fig. 5), or even syntrophy, where each network produces a waste product of value to the other. It seems reasonable to assume that cooperative systems would be more robust in the long run, but, while there has been some prior theoretical work (Hordijk et al., 2012; Vasas et al., 2012), more research is needed to evaluate the evolutionary consequences of passive or cooperative coexistence as a means of complexification.

Replacement also has the potential to systematically increase complexity, since it entails one life state replacing another. If complexity can be shown to correlate with competitive ability then we would have found an additional natural driver for increasing complexity in the pre-cellular stage. It is easy enough to propose certain chemical features that might be expected to confer a competitive advantage. For example, chemical systems whose food-harvesting steps have high affinity (e.g., low Km) might have a systematic advantage over systems with lower affinity for the same food compounds. Similarly, systems enriched for chemicals that can bind to the surface more strongly or have lower diffusion rates would also be expected to predominate over time (albeit while growing slower). However, while I can imagine several such chemical factors, none would obviously tend to correlate with complexity. Nonetheless, I will conjecture that there ought to be a general tendency for more complex autocatalytic states to replace less complex ones simply because the former have a superior ability to take-over (i.e., grow into) adjacent pixels.

Let us first consider a case in which a high complexity state, H, in pixel *p* interacts with a lower complexity state, L, in an adjacent pixel, *p'*, were H and L lie on the same diagonal through the origin in CC space. That is to say, the H and L states contain exactly the same species in the same proportion but at different absolute concentrations. Stipulating that L and H are both self-propagating in the absence of interactions with other life states, we know that each can convert an adjacent pixel from a state in equilibrium with the environment (with non-food species at concentration 0) into L or H, respectively. It stands to reason that if H can convert an adjacent pixel lacking each chemical $X_i$ (i.e., $[X_i] = 0$) to the H state, then it could certainly convert *p'*, which has the concentration of $X_i$ seen in the L state ($0 > [X_i]_L < [X_i]_H$). On the other hand, there is no reason to expect that L could convert

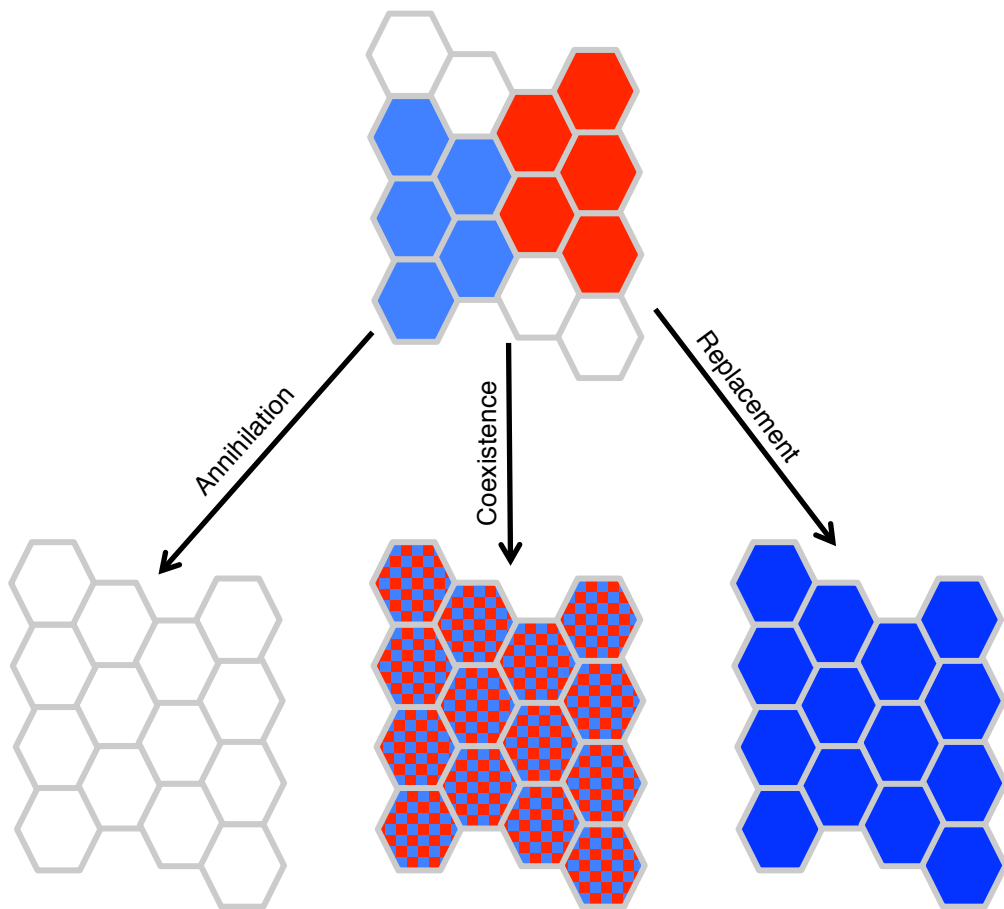

Fig. 6. Possible outcomes of interactions among two different life states, depicted in blue and red.

H to the L state. Assuming this is true for all non-food species we have reason to assume that H will replace L in any such case.

The same reasoning can be generalized to suggest that H will replace L in all cases where H and L are different growing life states and where all chemical species satisfy the inequality, $[X_i]_L \leq [X_i]_H$. For example, suppose that L and H have identical concentrations of all chemicals except for some chemical species that are part of the autocatalytic set acting in H (with a non-zero concentration) but absent from L. For example, let L be the autocatalytic set depicted in Fig. 3 and H be the one in Fig. 5, which also includes $C_4$ and $C_5$. It seems inescapable that H will replace L, meaning that the additional chemical species ($C_4$ and $C_5$) will be spread throughout the areas previously occupied by the L state.

These simplistic examples do not provide an explicit guide for cases in which some chemicals are at higher concentration in H and others are at higher concentration in L, even in the case that H has higher overall complexity (i.e., a larger Euclidean distance from the origin of CC space). That being said, I will conjecture that this should typically be so. Higher complexity chemical states can only succeed in growing – converting adjacent pixels to their high complexity state – by having high enough productivity to flood those adjacent pixels with molecules for each keystone species. This would seem to predict a general tendency for higher complexity states to replace lower complexity states, which could provide a general explanation for why lower entropy dissipative structures might tend to accumulate over evolutionary time. That being said formal mathematical models would be needed to validate this intuition and assess whether it holds just for chemical concentration or also applies to the internal energy term of chemical complexity.

Boundaries and cell formation

It is generally held that the cellular habit, the existence of populations of bounded vesicles that compete for representation in future generations, can make adaptive evolution more efficient. However, such an advantage for future evolution cannot be used as a direct explanation for the transition from an unbounded state, surface-associated living chemical ensembles, to bounded protocells. Nonetheless, I will now argue that adaptive evolution on surface life states, framed by the concept of chemical complexity, suggest a plausible evolutionary path to the cell.

One potential class of non-keystone chemicals gained during a living system's evolution are those that self-organize to form a membrane between the system and the overlying solution (Fig. 7 A-B). Such chemicals might first form as waste products (Nghe et al., 2015), but could then become enriched if their presence, directly or indirectly, promoted the rate of their own production in the system (for example by elevating the local concentration of reactants). If membrane-forming species inhibited diffusion of an essential food species into the system, or toxic waste product out of the system, they would be parasitic and could never spread through an entire living system. However, if the membrane allowed sufficient flux of food and waste chemicals for its own production, it could spread over an entire evolving system.

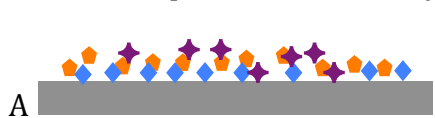
A

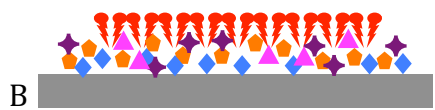
B

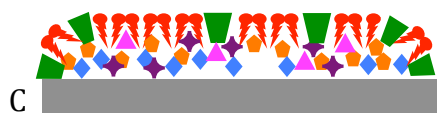
C

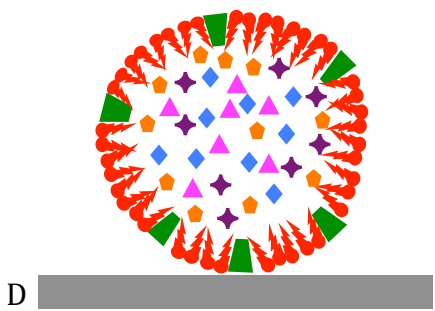
D

In the context of a surface receiving a flux of food species from an overlying solution, there are good reasons to expect that a membrane-bearing life state that arose, M, would tend to replace any non-membrane-bearing life state that it encountered, N. The logic behind this claim is that a membrane could only have come to characterize M if it the membrane were permeable to all species needed for M to survive, yet there is no a priori reason to think that this membrane would also allow rapid diffusion of the species needed for N to survive. As a result, should a patch of M come into contact with a patch of N, the membrane would be likely to render N inviable. This would disallow coexistence, but would also make it that much harder for N to outcompete M for shared food or attachment points. As a result, over time we might expect to see a higher and higher proportion of living systems to produce an overlying membrane.

Fig. 7. Scenario for the origin of membranes and the cell habit. A. The ancestral surface-associated life state is supposed to be composed of a set of keystone chemical species (purple, orange, blue) that are weakly attached, directly or indirectly, to the underlying mineral surface. B. An new amphiphilic chemical species (red) if formed and self-organizes to form an overlying membrane. The presence of a membrane allows the accumulation of novel chemical species (pink). C. The membrane is stabilized by the production of a further bridge species (green) that can associate with the membrane and/or the underlying mineral. D. Physical disruption can release a stable cell-like vesicle that would initially be capable of reestablishing the surface state through the actions of the bridge species (green).

In addition to providing a competitive advantage in pairwise interactions among life states, the addition of a selectively permeable membrane over a living system significantly changes that system's evolutionary potential. This follows because many dissolved chemical species could now become locally enriched between the membrane and the surface. Because the rate at which these chemicals are expected to equilibrate with the environment has slowed, the threshold productivity for autocatalytic systems is also lowered, effectively increasing the number of potentially viable metastable states. Thus, insofar as the vectors in CC space are scaled by diffusion, the addition of a membrane (or any other factor changing rates of diffusion for some or all species) could greatly change the expected dynamical behavior[16].

The formation of an overlying membrane requires some chemical connection, probably non-covalent, between the membrane and the underlying mineral surface, likely via bridging molecules (Fig. 7C). Such systems would be well suited to long dispersal of the life state to a new mineral surface. Physical agitation could generate a membrane vesicle that included all the keystone species of the life state as well as the bridging molecules that confer an affinity to the natal mineral on which the system evolved. As a result, such vesicles could serve as propagules, allowing the system to colonize all patches of natal mineral within some broad area (Fig. 7D). As argued previously (Baum, 2015), propagules of this sort provide a natural step towards the evolution of the cellular habit. At first the autocatalytic system might only have been metabolically active when interacting with the mineral, but selection would gradually favor variants that could grow and divide while dispersing in the liquid phase (Baum, 2015). Thus, selection for dispersal ability can be seen as the key evolutionary driver of the origin of the cellular habit.

The formation of cells changes the evolutionary dynamics by disadvantaging parasitic chemical species (Vasas et al., 2012). Whereas the spatial organization accorded by a two-dimensional surface will generally prevent parasites from taking over (Virgo et al., 2013), it also makes it very difficult to drive parasites to extinction. However, once life becomes compartmentalized into cells, selection for cellular viability and replication ability allow parasites (except those that, like modern viruses, can move horizontally among cells) to be exterminated.

---

[16] This resembles niche construction wherein evolutionary innovations of a taxon alter the relationship between genotypes and fitness and, hence, the topography of an adaptive landscape.

Even after compartmentalized living systems arise, we might expect it to take time for cells to function as autonomous, clonal lineages. The same tradeoff that drives modern life to experience varied frequencies of sexual reproduction would apply. Fusion would be favored because frequent sharing of chemical species among cells of the same evolving population would help prevent cell lineages from drifting away from their shared attractor. This follows because the average concentrations of chemicals in two cells would, on average, tend to be closer to the attractor than either cell will be individually.

Although fusion is favored as a means to reduce the drift of chemical concentrations, there are two major downsides to the promiscuous fusion of cells. First, fusion of cells that sit close to *different* attractors will tend to average the concentration of all chemical species and thereby move the fused cell out of the zone of attraction of either attractor. In other words, fusion of two living cells can immediately yield a non-living cell[17]. Second, fusion exposes each cell to the risk of picking-up a sexually transmitted parasitic species. Thus, the evolution of mechanisms to prevent fusion among cells or between cells and surface-associated relatives could be slow, but is likely to eventually arise, especially among distantly related cells.

It is worth highlighting that although the evolutionary transition to a population of individuated cells within little if any fusion would alter evolutionary dynamics (e.g., the frequency of parasites), the new system can still be accommodated within the dynamical systems view of life proposed here. In some sense, the CC space concept works better for cells than for uncompartmentalized, surface-associated systems because, in contrast to "pixels," cells are objectively individuated. Thus, it is reasonable to view the axes of cellular CC space as corresponding to the concentrations of each chemical species in a cell relative to that chemical's concentration in the environment (weighted by its internal energy). Thus, position in CC space well captures cellular complexity, at least until cells evolve subcellular organelles[18].

One way we can see that the CC space conception applies well to cellular life is to consider osmosis. CC space makes it explicit that there is a tendency for living systems to equilibrate by diffusion with the environment. Thus, it stands to reason that any chemical species that can freely cross a cell's membrane will tend to have the same internal and external concentration unless it is actively synthesized or degraded in the cell. Furthermore, insofar as the environment is enriched in species with low internal chemical energy, influx from the environment will tend to lower a cell's complexity. Thus, when water enters a cell and reduces the concentration of cytosolic components, it is indeed lowering the cells complexity. This pull towards the origin can be resisted transiently through kinetic blocks, such as mechanisms that reduce diffusion, but can only be sustained in the long run via the expenditure of energy. Thus, I would like to suggest that the CC space metaphor provides a general framework for thinking of cell complexity and provides a direct link to the principles explaining membrane transport and chemiosmosis.

---

[17] This is not unique to protocells with analog inheritance systems, such as those discussed here. F1 hybrid inviability is basically the same phenomenon.

[18] It should be noted that once cells acquire spatial substructures, such a macromolecular assembles or organelles, then a simple tally of the number of molecules of each type in a cell would underestimate its true complexity.

The evolutionary invention of autonomous cells entailed the gradual weaning of living systems from dependence on their natal mineral surface and the gradual acquisition of mechanisms to limit cell-surface and cell-cell fusion. One important consequence of the acquisition of population dynamics is that we can more easily model competition among cells that occupy different points in CC space. Multiple cells in a population might sit in the zone of attraction of the same metastable attractor, implying that they will tend to converge to the same point (the attractor) given enough time. These can be thought of as cells with the same genotype but with phenotypes that have been altered by environmental noise. Other cells, though, may have experienced a history of chemical mutation that results in them sitting in a different life state – one that might confer higher or lower fitness (i.e., potential for survival and reproduction). Thus, the essential aspects of conventional population genetics and individual level selection arise as soon as populations of bounded living systems arise.

Particulate inheritance and digital genetic encoding

At risk of taking the model too far, let me end by suggesting that the cell habit sets up conditions in which the evolution of a digital genetic system involving discrete heritable states encoded by informational polymer is highly plausible. Prior to cellularization, there is no reason to expect a genetic system like that seen in contemporary life. However, once cells arise, particulate inheritance should follow, characterized by particular molecular species being present as few (typically two copies) per cell prior to division, with mechanisms in place to assure that each daughter cell acquires an equal subset of these copies.

Although I have characterized CC space as having a continuous variable, concentration, on each axis, once life is cellularized axes are more accurately defined in terms of the number of molecules of each species relative to the number of molecules expected in a similar volume of the extracellular environment. For species present as many copies per cell the discrete nature of CC space is unimportant. However, provided cell volumes are low and the number of keystone species high then we should expect some chemical species to be at such low abundances that they will show quantized changes from generation to generation. Because random cell division has a high probability of leaving one daughter cell with no copies of these low-abundance keystone species, selection will clearly favor mechanisms that improve the reliability with which all daughter cells receive a full complement of keystone species. Such selection should yield molecular innovations that (a) postpone division until the rarest species reaches a sufficient number of copies per cell (Kamimura and Kaneko, 2010), and/or (b) ensure reliable physical segregation of low abundance keystone species to each daughter cell. Furthermore, once segregation mechanisms evolve, we might expect it to become used for a higher and higher proportion of the cell's low-abundance chemical species.

In principle, particulate inheritance could entail the passing on of multiple separate keystone species rather than any singular genome. However, many classes of polymers possess the right balance of covalent bonds and non-covalent interactions to be able to function both as catalysts and to encode digital genetic information (Baum and Lehman, 2017). Consequently, it is not hard to see how a subset of polymers might become specialized as repositories of genetic information (Takeuchi et al., 2017) and, if they did, how they might evolve into a single large and actively copied and segregated genome(Kamimura and Kaneko, 2010). Nonetheless, extensive additional work is needed to spell out all these steps and better understand whether a digital genetic code is an inevitable long term result of cellularization and if so, how this transition can or must happen.

Future work

In this paper I have proposed that viewing life as a localized, metastable chemical system that avoids equilibration with the environment through the use of environmentally provided matter and energy (i.e., food) helps clarify the origin and early evolution of life. Such a conception points to surfaces as the most likely milieu for self-propagating chemical systems (life) to originate, suggests that there may be a general tendency for surface-associated systems to complexify through neighborhood selection, lends credence to the idea that the cell habit is an expected innovation related to dispersal, and implies that particulate inheritance and digital genetic coding are likely innovations of any cellular life form that arises and evolves for long enough. However, while these verbal and conceptual arguments are tantalizing, I think the reader will agree that they need to be formalized more fully. For example, work is needed to understand conditions conducive to the spontaneous emergence of autocatalytic, self-propagating chemical systems, to clarify how such systems can move among multiple metastable equilibria, how competing states on a single surface will interact, and how surface associated states can generate membranes and cells. I therefore hope that the CC space conception of life will stimulate the generation of mathematical models that formally explore these and related question. Nonetheless, while I recognize that many readers will wish to suspend acceptance of this framework until rigorous, quantitative models have been developed, let me end by arguing that we should not let the lack for formal demonstration discourage us from making use of this conceptual framework to guide empirical research.

Historically, research into the origin of life has been premised on the idea that life cannot get going until there exists some entity capable of self-replication. Whether we envisage that first self-replicator as a protocell or a very talented RNA molecule, this perspective can only make us pessimistic about ever seeing the spontaneous origin of new life in the lab (Baum, 2015). Consequently, origin of life research has, primarily focused on the historical problem of explaining how certain distinctive chemical features of cellular life arose, especially nucleic acids and proteins. The ahistorical problem of how living systems in general originate and what features they *must* have has barely been studied empirically. After all, if you believe there is no hope of being able to generate a new living system in the lab how *could* you make progress except through theoretical modeling?

Taking the view, supported by the CC space metaphor, that any autocatalytic set capable of growing over a mineral surface is a living system changes the calculus greatly. If the great complexity of extant life is not a primordial feature but, rather, the consequence of complexity accreting over the last ~4 billion of years, it suddenly seems plausible that new living systems can be studied empirically. Such a perspective allows that the earliest stages of life might arise easily enough to be seen in the lifetimes of today's scientists. Indeed, the outlines of such a research program have already been proposed (Baum and Vetsigian, 2016) and are already being acted upon by several research groups, including my own. Thus, by conceptualizing life in the broader CC framework we may hope that a new generation of ahistorical origin of life research will begin, one that will finally help us appreciate that life is just a special kind of chemistry that has the *potential* to complexify over eons of time.

Acknowledgements
I thank many colleagues for useful discussion on these topics over the years, especially Domenico Bullara, Irving Epstein, Niles Lehman, Kalin Vetsigian, and Lena Vincent. I gratefully acknowledge funding from NSF (CHE-1624562), NASA (IDEAS16-0002), the University of Wisconsin-Madison Office of the Vice Chancellor for Research and Graduate Education with funding from the Wisconsin Alumni Research Foundation.

Competing interests
I declare no competing interests.